\documentclass{article}

\usepackage{amssymb,amsfonts,amsmath}
\usepackage{color,cancel}
\usepackage{graphicx}
\usepackage[normalem]{ulem}
\RequirePackage{lineno} 

\title{Sensing and decision-making in random search}

\author{Andrew M. Hein\thanks{Department of Biology, University of Florida, Gainesville, FL, \texttt{amhein@ufl.edu}} \and Scott A. McKinley\thanks{Department of Mathematics, University of Florida, Gainesville, FL \texttt{scott.mckinley@ufl.edu}}}

\begin{document}

\maketitle

\begin{abstract} While microscopic organisms can use gradient-based search to locate resources, this strategy can be poorly suited to the sensory signals available to macroscopic organisms.
We propose a framework that models search-decision making in cases where sensory signals are infrequent, subject to large fluctuations, and contain little directional information. 
Our approach simultaneously models an organism's intrinsic movement behavior (e.g. L\'evy walk) while allowing this behavior to be adjusted based on sensory data. We find that including even a simple model for signal response can dominate other features of random search and greatly improve search performance. In particular, we show that a lack of signal is not a lack of information. Searchers that receive no signal can quickly abandon target-poor regions. Such phenomena naturally give rise to the area-restricted search behavior exhibited by many searching organisms. 
\end{abstract}

Living organisms routinely 
locate resources in complex 
noisy environments. While gradient-based search provides a solution to this problem for microscopic organisms \cite{berg:1993}, the mechanisms for search decision-making in macroscopic organisms are not yet understood. This is because the sensory signals that macroscopic searchers encounter may be infrequent, subject to large fluctuations, and contain little directional information \cite{vergassola:2007}. To solve this problem, many species appear to perform area-restricted search: first searching large spaces 
for sub-regions that contain resources, then using sensory signals such as visual cues, vibrations, and chemical scents to precisely locate targets 
\cite{knell:2011, bell:1991}. For example, it has been proposed that some sharks 
adopt random movement strategies to locate resource patches \cite{humphries:2010}, but these species clearly use sensory cues to find nearby prey \cite{gardiner:2007,kalmijn:1982}.  
Efforts to study this process have focused either on large-scale random search behavior such as L\'evy walks in the absence of sensory signals \cite{viswanathan:1999, viswanathan:2008,bartumeus:2002,james:2011}, or on relatively small-scale search using readily available but noisy sensory cues \cite{vergassola:2007,barbieri:2011, torney:2009,balkovsky:2002}. 

These two components of biological search have generally been studied separately. By consequence, it is not clear how organisms combine large-scale resource localization behavior and small-scale sensory search to reliably find resources. Two aspects of this process are particularly obscure. First, it is unclear how organisms make the decision to transition back and forth between large-scale resource localization behavior and local search. Second, studies of large-scale random search have not typically considered the possibility that organisms use sensory data to inform search decisions. The degree to which such data might affect search performance 
is therefore unknown.
Here we present a mathematical framework that models search decision-making across large habitats containing regions of high and low signal availability.  The framework incorporates the idea that organisms adopt 
statistical movement strategies \cite{viswanathan:2008} but it allows such behavior to be modified by the collection and interpretation of noisy sensory cues. To explore how integrating random search and sensory signals 
affects search performance, we develop an individual-based model of a searching predator. We use the model to compare search performance of non-sensory predators that make search decisions using purely random strategies (L\'evy walk and a novel diffusive strategy), to signal-modulated predators that modify random strategies based on olfactory signals.

\section{Model development}

To study search decision-making, we consider an idealized model of a visual-olfactory predator in search of prey. We allow the predator to navigate through a large two-dimensional habitat in which prey are sparse and heterogeneously distributed.  
We assume prey emit a scent that can be detected by nearby predators. Searching predators will therefore sometimes be located in regions that are close to prey (tens to hundreds of body lengths) in which scent signals are available, but will also have to navigate through regions that are very far from prey (hundreds to thousands of body lengths) in which signals are absent. 

\begin{figure}[h!,width = \textwidth]
	\center
\includegraphics[width = 0.6\textwidth]{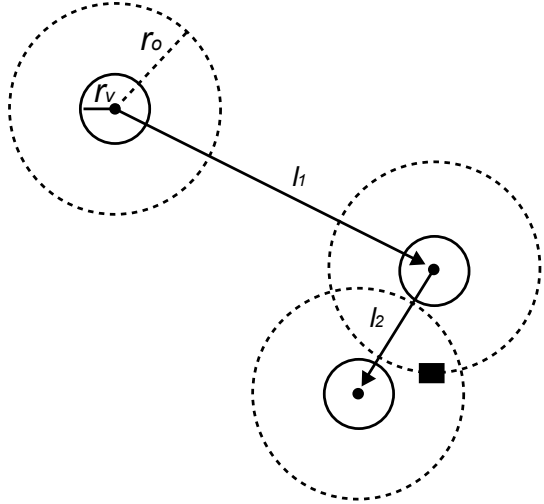}
\caption{Schematic of predator search. During the scanning phase of the search, the predator (black circle) detects prey (black square) within a radius of $r_v$ (solid inner circle) and detects scent signals emitted by prey within a radius of $r_o$ (dashed outer circle) at an average rate of $\geq 1$ per $\tau_o$ units of time. The predator then turns a random uniform angle between 0 and $2\pi$. During the movement phase, the predator moves a distance of $l$ units determined by its step length distribution and the $h$ scent signals detected in the last scanning phase.}
\label{schematics}
\end{figure}

Similar to previous approaches (e.g. \cite{lomholt:2008} 
), we assume predators search according to an intermittent process divided into two phases: a local scanning phase and a movement phase (Figure 1). Intermittent search is observed in a wide variety of organisms in nature \cite{kramer:2001}. 
During the scanning phase, if any prey are present within the predator's vision distance $r_v$ (Fig. 1, solid inner circle), the predator locates them with probability one. This reflects the high local acuity of vision. Signal-modulated predators also gather and process scent signals. The duration of the scanning phase is denoted $\tau_v$ and $\tau_o$ for non-sensory and signal-modulated predators respectively. $\tau_v$ includes the time needed to visually search a region of radius $r_v$ and reorient before taking another step, while $\tau_o$ includes the time taken to collect and process olfactory signals, visually search a region of radius $r_v$, and reorient before taking another step. The olfactory radius $r_o$ (Fig. 1, dashed outer circle) is defined as the distance at which the predator registers an average of one scent signal per scanning period $\tau_o$ (see below). We assume that each prey item emits scent at rate $\lambda$.
In the movement phase, the predator travels in a random uniform direction, a distance of $l$, at speed $v$. For non-sensory predators, the step length $l$ is drawn from some prescribed step length distribution $\pi(l)$, examples of which are described in the next subsection. Signal-modulated predators draw from a modified step length distribution defined below by equation \eqref{eq:bayes2}. During the movement phase, we assume that the predator is not capable of locating prey or detecting scent signals. Additionally, we assume that the predator only responds to its most recent scent signal information and does not store information about particular spatial locations. 

We study the limiting case in which sensory signals are rare, contain little information, and are not remembered by the searcher because this is the scenario in which random search strategies are often invoked. Our analysis thus evaluates the scenario in which sensory data are \emph{least} likely to yield improvement over non-sensory search. However, we point out that more sophisticated strategies are possible when searchers are capable of remembering past signal encounters or previously visited locations \cite{vergassola:2007,barbieri:2011,berg:1993}. 

\subsection{Non-sensory search}
It has been argued that organisms may rely on intrinsic movement strategies to encounter resources with unknown locations. In particular, the random foraging hypothesis holds that the movements of searching organisms can be described as random strategies (resulting in random walks), in which the particular properties of the strategy are altered through natural selection to optimize the rate of encounters with targets \cite{viswanathan:2008}. This hypothesis, which has been applied to searching organisms ranging from bees \cite{viswanathan:1999} to sharks \cite{sims:2008}, is typically invoked when it is not possible or practical for searchers to remember explicit spatial locations \cite{viswanathan:2008} and the typical distances between targets exceeds the sensory range of the searching organism \cite{humphries:2010}. The random search strategies that are typically studied do not involve collection or use of sensory data. One way to interpret random strategies is that they describe an organism's underlying tendency to move in a particular manner in the absence of strong stimuli.

When modeling search decision-making of organisms using random strategies, movement phases are typically determined by two randomly generated decisions: a step length and a turn angle. The details of these distributions determine the asymptotic properties of the search.
Random strategies are often categorized by this asymptotic behavior: diffusive behavior in which long-term mean-squared displacement (MSD) scales linearly with time, and superdiffusive behavior, including L\'evy walk behavior, in which MSD increases superlinearly with time. Depending on the details of the setting, studies have shown that either the diffusive \cite{benhamou:2007,bartumeus:2002} or superdiffusive \cite{bartumeus:2002,bartumeus:2008,james:2008, james:2011} strategies can be more effective.

We model predator movement using two random search strategies, one diffusive and one superdiffusive.  For both, we take the distribution of turn angles $\theta$ between successive steps to be iid $\theta \sim \mathrm{unif}(0, 2\pi)$ \cite{viswanathan:1999}. For the superdiffusive predator, we consider a L\'evy strategy, which draws step lengths from a Pareto distribution, $\pi_L (l) = (\alpha -1) l_m^{\alpha - 1} l^{-\alpha}$, with tail with parameter $\alpha$ and minimum step length $l_m$ (Fig. 2A solid red curve, $1 < \alpha \leq 3$ \cite{viswanathan:1999}). For the diffusive predator, we introduce a new step-length distribution which we call the \emph{true distance distribution} (TDD) $\pi_T(l)$ (Fig. 2A dashed blue curve).  
This strategy represents a greedy strategy wherein the predator 
selects step lengths from the probability distribution of the unknown distance to the nearest prey item (see Materials and Methods for further discussion; TDD given by equation \eqref{eq:tdd}). We chose this particular strategy because it is quite distinct from the L\'evy strategy (compare curves in Fig. 2A) and later serves to illustrate the strong homogenizing effect of signal-modulation on search behavior.
 
 \begin{figure}[h!,width = \textwidth]
	\center
\includegraphics[width = 0.3\textwidth]{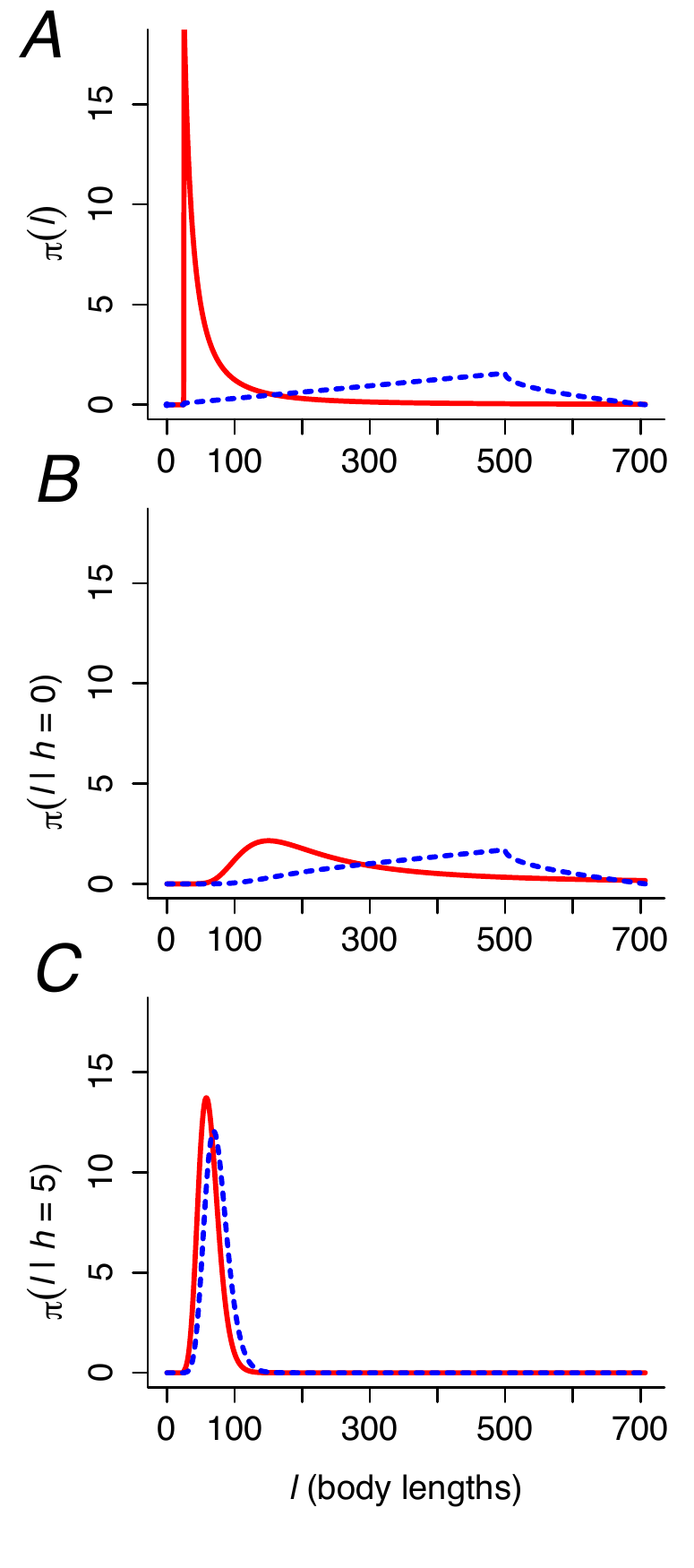}
\caption{Step length distributions before and after signal modulation. \textbf{A)} Step length distributions corresponding to L\'evy (solid red curve, $\alpha = 2, l_m = 50$ body lengths) and TDD (dashed blue curve) strategies prior to signal-modulation, \textbf{B)} after signal modulation when $h = 0$, and \textbf{C)} after signal modulation when $h = 5$. The TDD strategy was computed assuming prey are separated by 1000 predator body lengths (see Materials and Methods). Signal modulated strategies were computed assuming the following: $a = 1$ predator body length, $l_m = 50a$, $r_o = 250a$, $r_v = 50a$, and $\lambda_a = 100$ units of scent per $\tau_o$(see text for description of parameters). } 
\label{sl_distributions}
\end{figure}

\subsection{Using sensory signals to inform search decisions}

We hypothesize that predators incorporate sensory signals into search decision making through two steps. First, we assume that the predator has a means of interpreting a particular signal observation to yield information on the likely distances to the nearest prey. This feature allows a predator to sample its environment and determine whether prey are likely to be nearby or far away. Second, we propose that the predator has a means of modifying its intrinsic tendency to move in a particular way (represented by a non-sensory search strategy $\pi(l)$) with the information gained from sensory signals. This feature allows the information gained by receiving and interpreting signal data to be exploited to guide movement decisions. Below we develop a model for incorporating olfactory signals into search decision-making, but note that this framework (and equation \eqref{eq:bayes2}) could be modified to model responses to other types of sensory cues.

In keeping with recently proposed models of olfactory search \cite{vergassola:2007,barbieri:2011}, a scent signal datum results when the predator collects olfactory information for $\tau_o$ units of time and encounters $H \in \{0, 1, 2, \ldots \}$ detectable units of scent. 
The problem of how to locate prey most effectively involves optimizing the use of signal data to guide movement decisions. In this scenario, two distinct uncertainties can influence predator movement decisions. The first is the uncertain distance to the target, characterized by the probability distribution $\nu$; and the second is the optimal step length distribution $\pi$ given $\nu$.  We assert that correctly determining $\pi$ from $\nu$ remains an unsolved, and perhaps intractable, problem. Ideally, given sensory data $H$, one would calculate a Bayesian posterior for the distance distribution $\nu|_H$ and then take the next step from the associated optimal step length distribution $\pi|_H$. However, in the absence of a clear understanding of how to determine optimal step length from a given distance distribution, we apply the Bayesian update to the step length distribution $\pi$ itself:
\begin{equation}
 \pi(l|H = h) = \frac{P(H = h| l ) \, \pi(l)}{\int_0^{\infty} P(H = h | l) \, \pi(l)\, dl}  
\label{eq:bayes2}
\end{equation}
This approximation to the ideal strategy yields significant improvement in expected search time (see Materials and Methods for further elaboration). 

Equation \eqref{eq:bayes2} combines an underlying non-sensory step length distribution [$\pi(l)$] such as those described in the Non-sensory Search section with a likelihood [$P(H=h| l )$], which represents the predator's perceived probability of registering $h$ scent signals during a scanning phase, given that the nearest prey item is $l$ units away. Thus, for any particular value of $h$, the predator's movements will reflect both its underlying tendency to move in a particular manner, and the information about the relative location of prey gleaned by encountering $h$ scent signals. 

Figure 2B-C shows the signal-modulation of both the L\'evy and TDD step length distributions for $h = 0$ and $h = 5$. When $h = 5$ (Fig. 2C) the probability that the source is many body lengths away is low and the predator is biased toward making local moves. When $h=0$ (Fig. 2B)
, the likelihood is near zero beyond the predator's vision radius and the predator is more inclined to make long moves. Nearby sites can remain unexplored with little risk that prey that might be present there. 
In the cases of $h = 0$ and $h = 5$, Figure 2 shows that signal-modulation causes the two non-sensory strategies to become much more similar. The use of information reduces the very distinct tendencies represented by these two strategies. 

\subsection{Interpreting scent signals}
 We assume the predator has evolved a means to estimate (or intuit) the probability of registering $h$ scent detections in $\tau_o$ units of time, as a function of its distance to the nearest prey.
This amounts to being able to estimate the likelihood $P(H=h|l)$, which depends on the process of scent propagation. 

At macroscopic scales, turbulent fluctuations in fluid velocity lead to large local fluctuations in scent concentration \cite{shraiman:2000}. We model scent arrival under these conditions as packets that appear at the prey position $\mathbf{x_0}$ according to a Poisson arrival process and then move as a Brownian motion. From the predator's perspective, this is equivalent to encountering a random number of units of scent, $H \sim \text{Pois}(\tau_o R(|\mathbf{x} - \mathbf{x_o}|))$, at its location $\mathbf{x}$ during a scanning phase of length $\tau_o$, where $R$ is the rate of scent arrival defined by equation \eqref{eq:pde_soln} (see Materials and Methods). Denoting $l = |\mathbf{x} - \mathbf{x_0}|$, under these assumptions, the likelihood of $h$ encounters is $P(H = h| l) = [\tau_o R(l)]^h e^{-\tau_o R(l)}/h! $. 

Our approach assumes that predators have acquired a means of estimating the likelihood either through evolution or through learning. Yet, the likelihood depends on values of several physical parameters (e.g. the rate at which detectable patches of scent decay) that may be difficult for a predator to infer from measurements of its physical environment. We therefore take a qualitative view in prescribing the parameters of scent propagation. The most important qualitative feature 
is the length scale $r_o$, which corresponds to the distance at which a predator will register on average one unit of scent per scanning period $\tau_o$.  Heuristically, this is the distance at which the predator is likely to detect a faint, yet non-trivial scent. 
A second qualitative restriction is the expected number of encounters per unit $\tau_o$ at a distance of one body length from the prey $\lambda_a$.
These two specifications constrain the problem of scent propagation so that the physical parameters can be determined. 

The quantities $r_o$ and $\lambda_a$ are much more readily measurable by a searching organism than are the explicit parameters in equation \eqref{eq:pde_soln}. It thus seems likely that these quantities may constitute part of an organism's \textquotedblleft olfactory search image" \cite{cross:2010}, and may serve as the direct measurements useful for  reinforcement learning.   
Responses to these measurements may also be canalized through natural selection, in a sense \textit{tuning} a predator to a particular type of prey.

\subsection{No signal does not imply no information}
 When searching large regions for sparsely distributed prey, predators will frequently be too far from prey to receive scent signals. It is therefore particularly interesting to determine how non-sensory strategies are modified when the searcher receives no signal. When scent signals can only be detected very near prey (i.e. $r_o/r_v$ is small), sampling and not detecting a scent yields little information about the relative location of prey ($P(H = 0|l)$ approximately constant, Fig. 3 black curve). In this case $\pi(l|H = h) \approx \pi(l)$ and the non-sensory strategy is barely altered. However, when scent signals can be detected far from prey ($r_o/r_v$ large), not encountering a scent indicates that prey are not nearby (i.e. $P (H = 0|l)$ is near zero in the vicinity of predator, Fig. 3 dark and light blue curves). Thus, when the length scale traveled by prey scent is sufficiently large, receiving no signal can still result in a change in the non-sensory search strategy. For example, Figure 2A shows that the L\'evy strategy has a high probability of taking small steps between re-orientations. When there is no signal, signal-modulation eliminates the likelihood of wasted local steps (Fig. 2B). Figure S4 shows that even by responding only to no-signal events and ignoring cases in which $h>0$, a signal-modulated L\'evy searcher can substantially decrease mean search time.

\begin{figure}[h!,width = \textwidth]
	\center
\includegraphics[width = 0.6\textwidth]{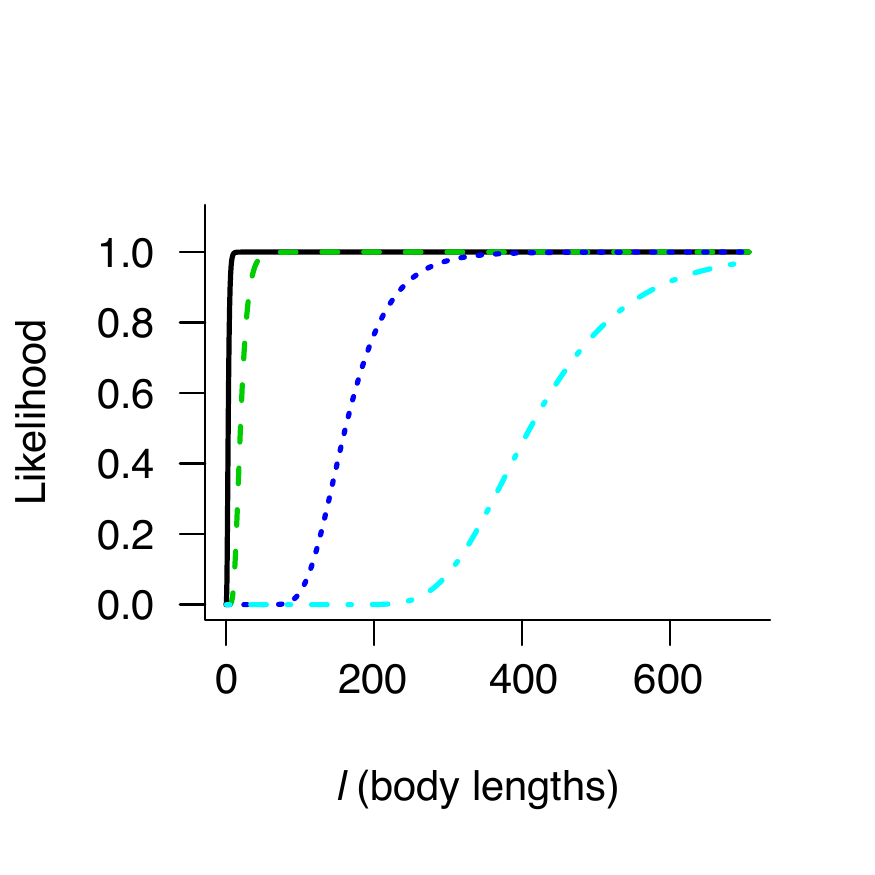}
\caption{Likelihoods ($P(H = h|l)$) resulting from receiving $h = 0$ scent signals in a particular scanning period. When the ratio of olfactory to vision radius is small (solid black curve: $r_o/r_v = 0.25$; dashed green curve: $r_o/r_v = 1$), encountering zero units of scent reduces the likelihood only very near the predator. As $r_o/r_v$ increases, the likelihood becomes small for many body lengths from the predator (dotted dark blue curve: $r_o/r_v = 5$; dot-dashed light blue curve: $r_o/r_v = 10$). All other parameters as in Fig. 2.}
\label{likelihoods}
\end{figure}

\section{Simulation results}

Mean search times of simulated non-sensory and signal modulated predators are shown in Figure 4A. Non-sensory predators relying on the L\'evy strategy with optimal tail parameter ($\alpha = 3$, Fig. 4A, red solid line) have lower mean search times than predators drawing step lengths from the TDD strategy  (Fig.  4A, blue dashed line). When the olfactory radius $r_o$ is similar to the vision radius $r_v$, signal-modulated L\'evy and TDD strategies exhibit mean search times comparable to their nonsensory antecedents (Fig.  4A, circles; red circles represent results from signal-modulated L\'evy with optimal $\alpha$, where optimal $\alpha$ was in the range 2.6-3.0 for all $r_o/r_v$). However, as the ratio $r_o / r_v$ increases beyond one, the mean search times of the signal modulated strategies drop dramatically. This demonstrates that searching predators can markedly improve search performance by incorporating signal data, so long as olfactory signals can be detected at a distance that is greater than the predator's vision distance. 

\begin{figure}[h!,width = \textwidth]
	\center
\includegraphics[width = 0.4\textwidth]{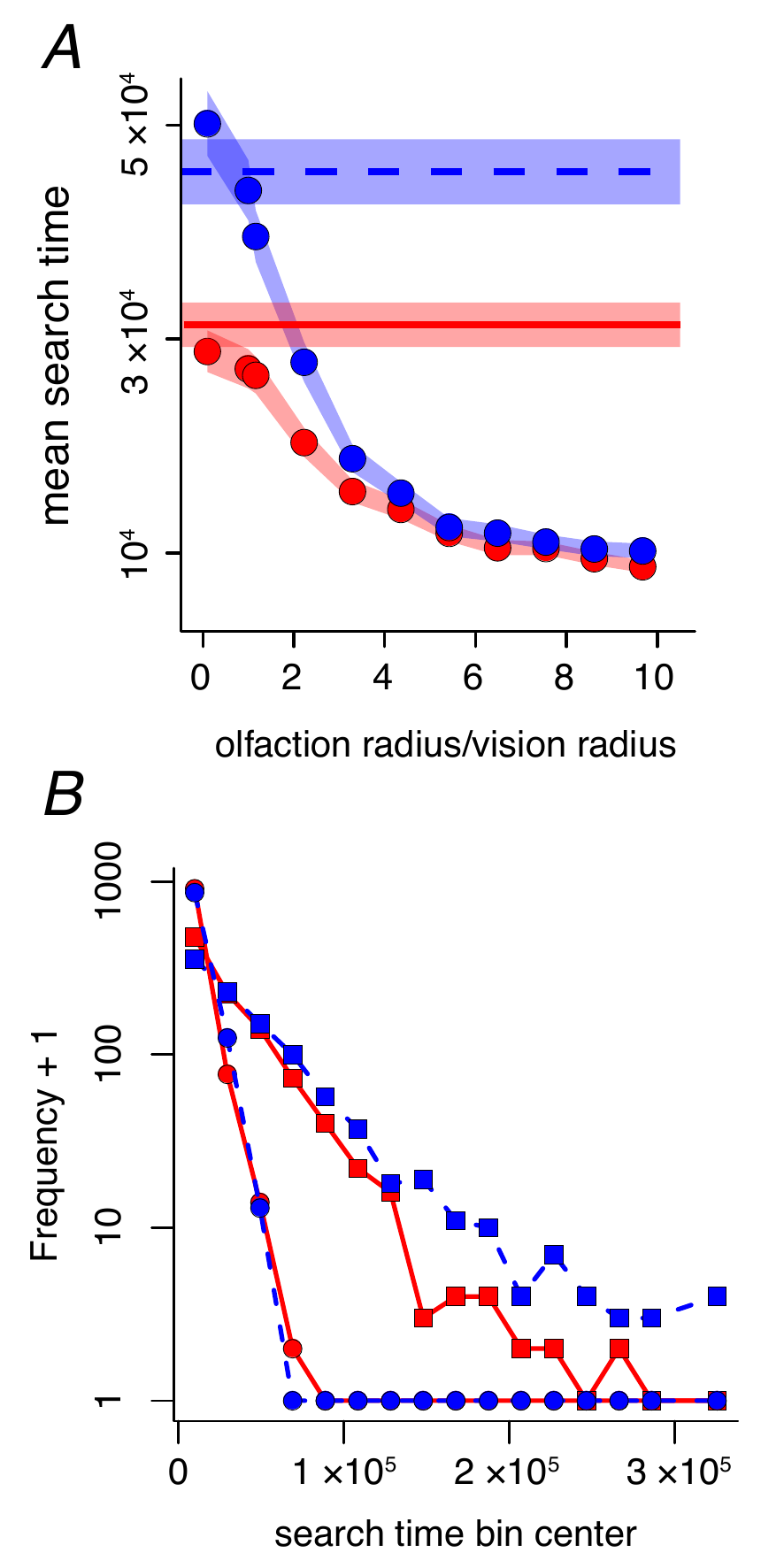}
\caption{Predator search times. \textbf{A)} Mean search time as a function of the ratio of the olfactory radius ($r_o$) to vision radius ($r_v$). Lines (non-sensory strategies) and points (signal-modulated strategies) each represent mean search time of 1000 replicate simulations. Bands represent $\pm2$ SEM. The following parameters values were used: $a = 1$, $r_v = l_m = 50a$, $\tau_v = 1$ s, $\tau_o = 30$ s, mean inter-target distance $L = 1000a$, and $\lambda_a = 100$ units of scent per $\tau_o$. \textbf{B)} Empirical search time distribution of L\'evy (red squares), TDD (blue squares), signal-modulated L\'evy (red circles), and signal-modulated TDD (blue circles) strategies. In the case of the signal-modulated strategies, frequencies are shown for the $r_o$ value resulting in the lowest mean search time. Note the large number of searches resulting in long search times for the non-sensory strategies.}
\label{search_times}
\end{figure}

The improvement in performance of the signal-modulated strategies is primarily due to the reduction in the number of searches that result in large search times. Figure 4B shows that the tails of the search time distributions for the signal-modulated strategies (Fig.  4B, circles) decay roughly exponentially at a rate that is much faster than the decay rate of the  non-sensory L\'evy and TDD strategies (Fig.  4B, squares). Notably, the risk of searches exceeding any particular threshold time is much greater for the non-sensory strategies.

\begin{figure}[h!,width = \textwidth]
	\center
\includegraphics[width = 0.8\textwidth]{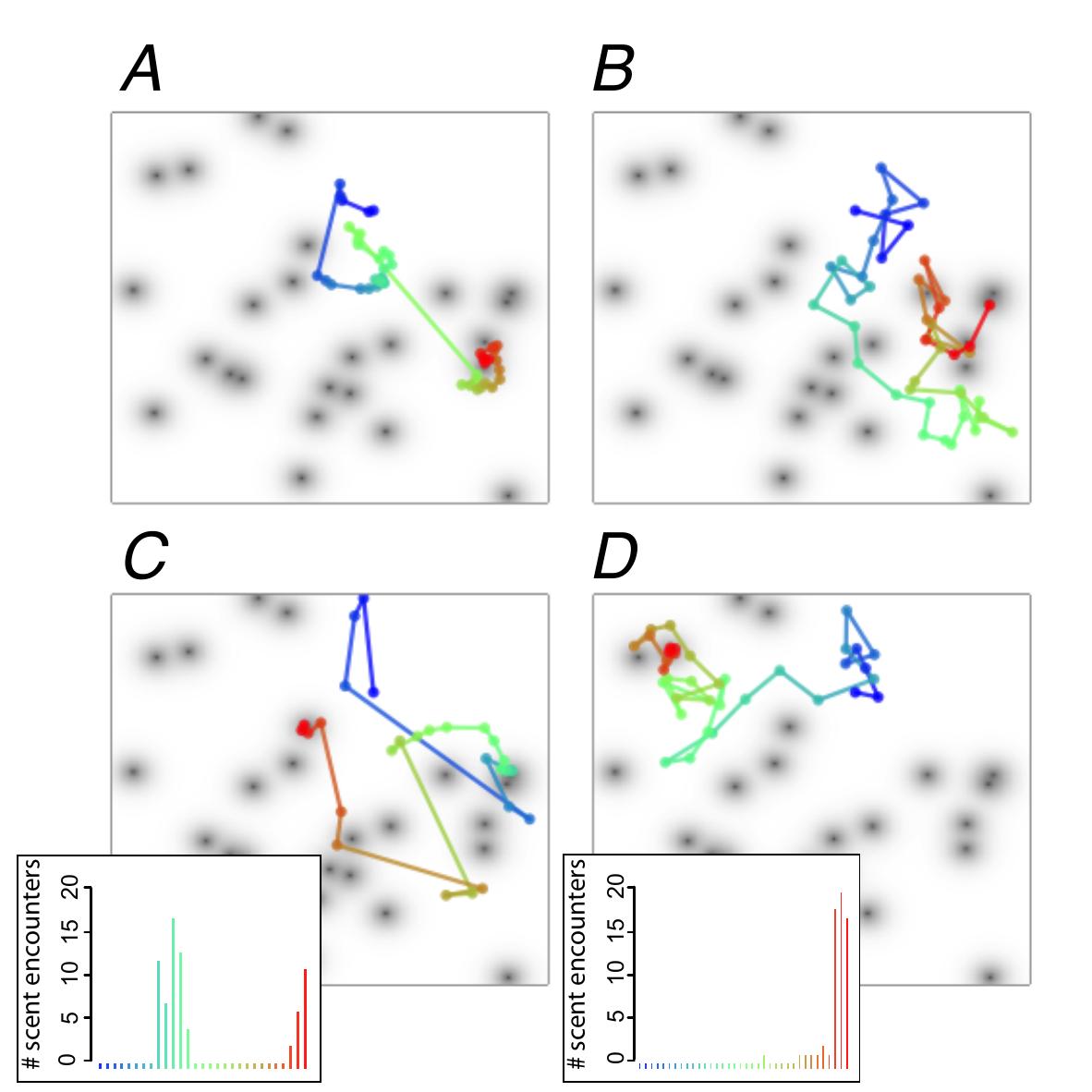}
\caption{Typical search paths through a scent field with $\log_{10}$(1 + mean number of scent encounters per unit $\tau_o$) indicated by grayscale (darker grey denotes more encounters, white indicates mean number of encounters $\approx 0$). Paths for \textbf{A)} L\'evy, \textbf{B)} TDD, \textbf{C)} signal-modulated L\'evy, and \textbf{D)} signal-modulated TDD are shown. Color scale of path changes from blue to red with increasing time. Inset panels in \textbf{C} and \textbf{D} show the number of hits received during each scanning period with colors corresponding to colors in search paths. Ratio of olfactory to vision radius was set equal to 4; all other parameters as in Fig. 4A.}
\label{search_paths}
\end{figure}

Figure 5 shows typical search patterns of predators using each of the four search strategies in the regime where $r_o > r_v$. While non-sensory predators exhibit similar behavior in regions near and far from the prey (Fig. 5A,B, lines), signal-modulated predators exhibit an increased tendency to make short exploratory steps in the vicinity of the prey (Fig. 5C,D, lines). Moreover, the pure L\'evy strategy over-samples regions that are far from prey (Fig. 5A), while the TDD strategy tends to take excursions that are too large to allow it to remain near a target for long (Fig. 5B). The inset panels in Figure 5C and 5D show that the number of signals received in scanning phases is typically zero, with signals of greater than zero only occurring when the predator is relatively close to prey. When far from prey, signal-modulated predators exhibit steps that are long enough to move out of the region of low probability directly surrounding the predator and avoid wasting time in local exploration.

\section{Discussion} 

Using sensory data to govern transitions in search behavior
can greatly improve search performance, even when sampling for sensory signals typically results in the collection of ``no signal". Indeed, under a wide range of conditions, a searching organism that receives no signal can infer that its target is far away and can respond by moving a relatively long distance before turning again. When a signal is detected, the searcher can respond by remaining in a local region. This type of signal-dependent response may underlie the area-restricted search behavior exhibited by a wide variety of species such as the desert isopod, the insect parasite \textit{Trichogramma evanescens}, and the Wandering Albatross \cite{knell:2011,bell:1991,weimerskirch:2007}.

Figure 5 shows that searching predators move from regions that lack prey through regions that contain prey and back again. Signal-modulated predators transition between making many short exploratory moves and making long moves 
depending on whether prey are likely to be nearby. The non-sensory search strategies do not do this; while they allow for different types of movements (e.g. long and short movements), these movements are not matched to the distance between the predator and the nearest prey. The L\'evy strategy provides a good example of this. This strategy favors relatively short steps (Fig. 2A) and occasional long steps. Short steps are effective when prey are nearby, but result in extensive exploration of regions that do contain prey (Fig. 5A). The signal-modulated L\'evy strategy, on the other hand, appropriately takes long steps when far from prey, and short steps when prey are nearby (Fig. 5C). It is worth noting that the signal-modulated searchers exhibit similar behavior to their non-sensory counterparts when $r_o/r_v$ is small. In this regime the L\'evy strategy outperforms the TDD strategy (Fig. 4A). This suggests that the distinction between different types of random strategies may still be important when resources can only be detected at very short distances. In such cases signal encounters will be exceedingly rare, and searchers learn little about the relative location of targets when the fail to receive a signal. 

Signal-modulated predators exhibit large scale relocation movements interspersed with periods of local search. This qualitative pattern has been observed in foraging marine fish and reptiles \cite{sims:2008}, ants in search of colony-mates \cite{franks:2010}, and many other searching organisms \cite{viswanathan:2008}. Our results suggest that this behavior can 
emerge naturally when searchers make different types of movement decisions in response to resource cues. 
This adds to the growing evidence that apparently heavy-tailed step length distributions may result from a variety of behaviors that include behavioral response to environmental cues \cite{boyer:2004} and mixing of different types of movement behavior \cite{benhamou:2007,petrovskii:2011}. 

Past studies have shown that the relative performance of random searchers depends heavily on the details of the search behavior, with different types of behavior (e.g. L\'evy vs. Brownian strategies) being more or less effective depending on the environment \cite{bartumeus:2002, james:2011}. Our model shows that modifying search decisions using sensory data can reduce the strong dependence of search performance on the particular type of random strategy taken. 
This result has interesting implications for the evolution of search behavior. For example, it has been suggested that random search behavior may emerge because underlying neurological activity predisposes searching organisms to make strong directional re-orientations at random intervals \cite{bartumeus:2008}. 
Our results suggest that signal-modulation may allow for a search response that is robust to fluctuations or differences in underlying tendencies to make one particular type of random movement or another. Our framework provides a means of studying effects of evolutionary changes in both the underlying random movement strategies and the manner in which these tendencies are modified by sensory information.

\section{Materials and Methods}
\subsection{Scent propagation} To see how $R(l)$ depends on the distance between predator and prey, let $u(\mathbf{x})$ represent the mean concentration of scent at predator position $\mathbf{x}$ emitted by a prey item located at position $\mathbf{x_0}$. An expression for the steady-state diffusion process without advection is given by
\begin{equation}
0 = D \Delta u (\mathbf{x})- \mu u(\mathbf{x}) + \lambda \delta (\mathbf{x}_0)
\label{eq:sspde}
\end{equation}
where $D$ represents the combined molecular and turbulent diffusivity ($ m^2 s^{-1} $),  $\mu$ represents the rate of dissolution of scent patches  ($s^{-1}$), and $\lambda$ represents the rate of scent emission at the prey ($s^{-1}$). In two dimensions, the rate of scent patch encounters by a predator of linear size $a$ located at  $\mathbf{x}$ is given by $R(l) = \frac{2\pi D}{-\ln(a\psi)} u(l)$ where $\psi = \sqrt{\frac{\mu}{D}}$ \cite{vergassola:2007}. This implies

\begin{equation} 
R(l) = 2\frac{\lambda K_0(\psi l)}{-\pi \psi \ln(\psi a)}
\label{eq:pde_soln}
\end{equation}

where $K_0$ represents a modified Bessel function of the second kind.
\subsection{Simulation details} The Supplementary Information (SI) Appendix (including figures S1-S3) provides details on the robustness of results to changes in model parameters. For each of the four search strategies (L\'evy, TDD, signal-modulated L\'evy, and signal-modulated TDD), we performed simulations in which predators explored a periodic environment with 100 randomly placed prey. The rate at which scent patches were encountered is given by equation (\ref{eq:pde_soln}) summed over all prey. To determine $h$ in any particular scanning phase, we computed the rate of signal encounters based on the true distance between the predator and targets and then generated $h$ as a deviate from a Poisson distribution with mean equal to the product of the encounter rate and the olfaction time, $\tau_o$. Predators were assumed to travel at a constant speed of one body length per unit time. Environments were constructed so that prey density had a mean of 1 prey per $10^6$ squared body lengths, a realistic low density for prey (see SI Appendix). However, the qualitative results shown above hold for lower prey densities as well (see figure S1 in SI Appendix). In the case of the L\'evy strategies, we repeated simulations across a range of alpha values from $\alpha = 1.2$ to $\alpha = 3$. 

In each simulation, the searcher was positioned at a random location and allowed to move through the environment until it encountered a prey item. In each simulation, all prey items were randomly positioned according to a Poisson point process with the mean distance between targets ($L = 1000$ searcher body lengths in the simulations shown in Figs. 4-5) chosen to achieve the desired density. For each strategy, we performed 1000 simulations 
and recorded the time until first prey encounter in each simulation. 

\subsection{True distance distribution (TDD) and a comment on the use of Bayes rule} 

The TDD is first calculated assuming the search region is the domain $[0,L]^2$ with periodic boundary conditions. Without loss of generality, suppose that a target is located at the position $(L/2,L/2)$ and the initial position of the searcher is chosen uniformly from the domain. The distribution of the distance from the initial position of the searcher to the target is then given by
\begin{equation} \label{eq:tdd}
\nu(l) = 2\pi l - \mathbf{1}_{\{l > L\}} 8l\, \arccos(\frac{L}{l})
\end{equation}
where $\mathbf{1}_{\{l > L\}}$ is an indicator function taking the value of zero for $l < L$ and 1 otherwise. 

In equation \eqref{eq:bayes2} we introduced a modification of a default non-sensory step length distribution that improves search performance by incorporating signal data.  The modification has the form of a Bayesian posterior distribution, but strictly speaking, this is not an implementation of Bayes' Rule. A more probabilistically rigorous approach to incorporating signal data would be the following. After conducting an olfactory scan, an ideal predator would use the TDD as a prior to compute a Bayesian posterior distribution $\nu|_H$ for the distance to the target:
\begin{equation*}
	\nu(l|H = h) = \frac{P(H = h| l ) \, \nu(l)}{\int_0^{\infty} P(H = h | l) \, \nu(l)\, dl} 
\end{equation*}
where the likelihood is computed as described above. Completing the ideal strategy hinges on whether it is possible to characterize an optimal step length distribution for a given posterior. One might try to pose this as a variational problem.  Let $\mathcal{P}$ denote the space of all probability densities on $\mathbb{R}_+$ then a signal-modulation strategy can be defined in terms of a functional $\Phi: \mathcal{P} \to \mathcal{P}$.  So, using this notation, the two functionals studied in this paper are $\Phi_{TDD}$ which is simply the identity functional and $\Phi_{Levy}$ which satisfies $\Phi_{Levy} (\nu) \equiv \pi_L$ for all $\nu$. For an appropriate set of strategies $\Xi$, one seeks an optimal strategy, $\Phi_* = \text{argmin}_{\Phi \in \Xi} \{E[\tau_{\Phi}]\}$ where $\tau_\Phi$ is the random hitting time of the target by a signal-modulated searcher using strategy $\Phi$.\section{acknowledgments}
AMH was supported by a National Science Foundation Graduate Research Fellowship under Grant No. DGE-0802270. We thank J. Gillooly, E. Kriminger, and A. Brockmeier for helpful discussion and comments.


\appendix




\renewcommand{\thefigure}{S\arabic{figure}}
\setcounter{figure}{0}

\section{Robustness of results to search conditions}

\subsection{Target density} 
\begin{figure}[h!,width = \textwidth]
	\center
\includegraphics[width = 0.6\textwidth]{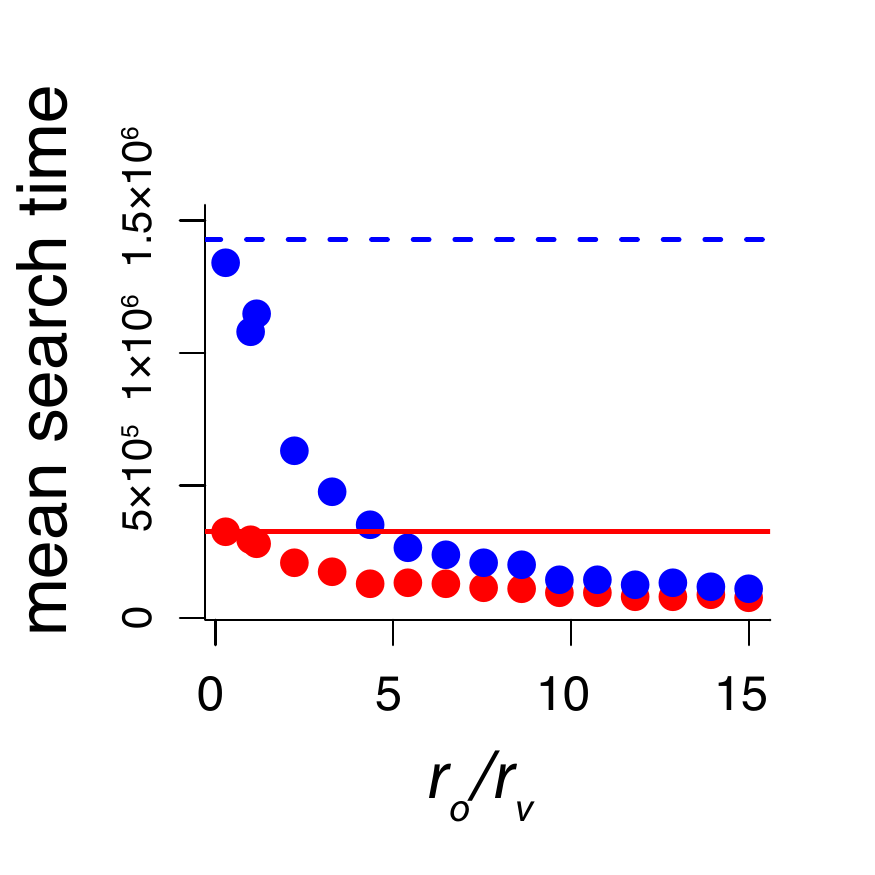}
\caption[Figure S1]{Mean search times for non-sensory L\'evy (red line), TDD (blue line), and signal-modulated L\'evy (red circles) and signal-modulated TDD (blue circles) predators. 200 replicate simulations were performed for each combination of strategy $\times$ $r_o/r_v$. The following parameters values were used: $a = 1$, $r_v = l_m = 50a$, mean inter-target distance $L = 3162a$, $\tau_v = 1$ s, $\tau_o = 30$ s, and $\lambda_a = 100$ units of scent per $\tau_o$.}
\label{fig:S1}
\end{figure}

In the simulations presented in the main text, we assume target density is one prey per $10^6$ square predator body lengths. This density is a realistic low prey density based on field estimates of prey densities for a variety of predators (e.g. \cite{karanth:2004,oli:1994},\cite{carbone:2002} and references therein). However, to determine whether our results hold at even lower prey densities, we repeated simulations after decreasing prey density by an order of magnitude (i.e. one prey per  $10^7$  square predator body lengths). Results from these low density simulations are shown in Figure \ref{fig:S1}. As in Figure 4A in the main text, mean search times of the signal-modulated L\'evy and signal-modulated TDD strategies decrease rapidly as the ratio of the olfactory radius to the vision radius ($r_o/r_v$)  increases above one. Moreover, these two strategies exhibit similar performance for large $r_o/r_v$ as in the results shown in the main text for higher prey density.

\subsection{Signal emission rate} 
\begin{figure}[h!,width = \textwidth]
\center
\includegraphics[width = 0.6\textwidth]{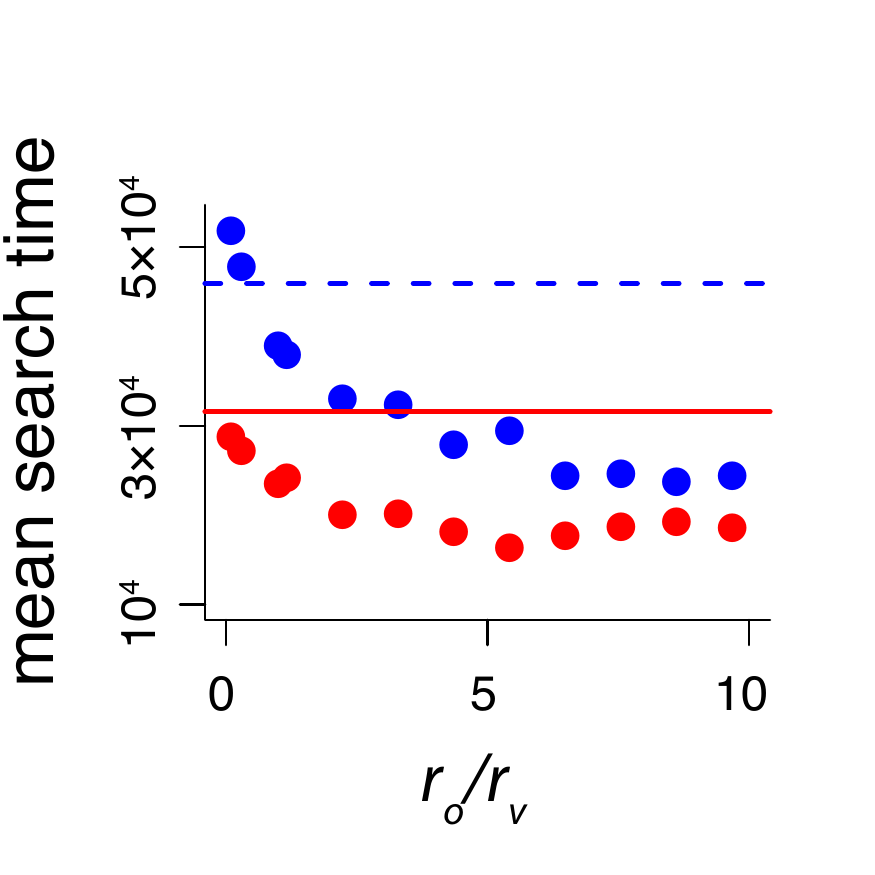}

\caption[Figure S2]{Mean search time with reduced rate of scent emission. Symbols as in Fig. S1. The following parameters values were used: $a = 1$, $r_v = l_m = 50a$, mean inter-target distance $L = 1000a$, $\tau_v = 1$ s, $\tau_o = 30$ s, and $\lambda_a = 10$ units of scent per $\tau_o$. Each point represents mean of 200 replicate simulations.}
\label{fig:S2}
\end{figure}

Simulations presented in the main text were conducted assuming the mean number of scent encounters per $\tau_o$ units of time was equal to 100 at a distance of one predator body length from a prey item (i.e. $\lambda_a = 100$). To determine whether results were qualitatively similar for lower emission rates, we repeated simulations after reducing $\lambda_a$ to 10 encounters per $\tau_o$ units of time. Results are consistent with those presented in the main text (Fig. \ref{fig:S2}). Mean search times of signal-modulated L\'evy and TDD predators decrease with increasing $r_o/r_v$. Search times of these two strategies also become more similar for large $r_o/r_v$. 

\subsection{Variation in predator scanning times} 
\begin{figure}[h!,width = \textwidth]
	\center

\includegraphics[width = 0.8\textwidth]{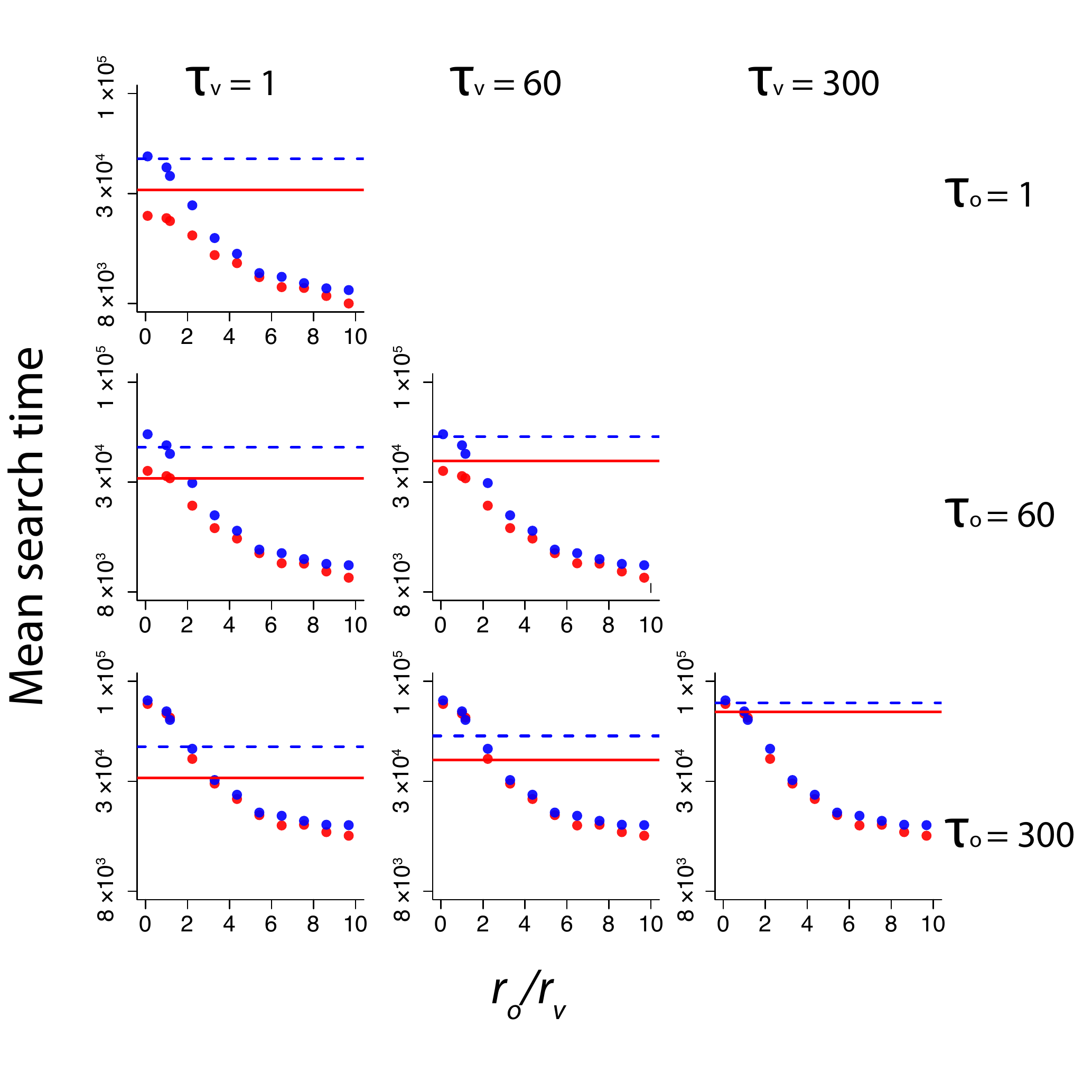}

\caption[Figure S3]{ Plot matrix showing lack of dependence of results on values of the $\tau_o$ and $\tau_v$ parameters. Symbols as in Fig. \ref{fig:S1}. Panels represent different combinations of $\tau_v$ and $\tau_o$ parameters ranging from 1 to 300. The following parameters values were used: $a = 1$, $r_v = l_m = 50a$, mean inter-target distance $L = 1000a$, and $\lambda_a = 100$ units of scent per $\tau_o$. Each point represents mean of 1000 replicate simulations.}
\label{fig:S3}
\end{figure}

Between successive steps non-sensory predators pause for $\tau_v$ units of time before taking another movement step, whereas signal-modulated predators pause for $\tau_o$ units of time. Typical pause durations between successive movements of a wide variety of animals in the field indicate that pause durations typically range from $\approx$ 1 s to $\approx$ 60 s \cite{kramer:2001}. Here we explore the robustness of the qualitative patterns shown in the main text to changes in the duration of the scanning phase for both signal-modulated and non-sensory predators. Scanning times may affect the relative performance of search strategies because some strategies (e.g. non-sensory L\'evy) pause more frequently than others. Moreover, differences between $\tau_v$ and $\tau_o$ determine the relative amounts of time spent scanning by non-sensory and signal-modulated predators. Figure \ref{fig:S3} shows mean search time as a function of $r_o/r_v$ for a range of values of $\tau_o$ and $\tau_v$. In all panels, mean search time decreases with increasing $r_o/r_v$ and mean search times of signal-modulated L\'evy and TDD are substantially shorter than mean search times of the non-sensory strategies for at least some range of $r_o/r_v$. It is worth noting that the relative performance of the L\'evy strategies versus the TDD strategies does depend on the absolute value of $\tau_v$ and $\tau_o$. This is because L\'evy strategies tend to go into the scanning phase more often and search times of these strategies therefore depend more strongly on scanning times.

\section{The role of \textquotedblleft no-signal" events}
In the main text, we discuss the potential importance of no-signal events, in which the searching predator samples for olfactory signals and receives zero signal (i.e. $h = 0$). Figures 2B and 3 in the main text show how the behavior of signal-modulated predators can be altered when $h = 0$, depending on the length scale at which olfactory signals are transmitted. The effect of zero signal on the L\'evy strategy is particularly strong because the probability of making relatively short steps is large, but the likelihood that a source is nearby given that $h = 0$ is low. Because of this property, this strategy is much more strongly influenced by receiving no signal than the TDD strategy (compare Fig. 2A and 2B in the main text). Another common model used in simulations of animal movement, the exponential step length distribution, also has this property. 

\begin{figure}[h!,width = 100mm, height = 100mm]
\center
\includegraphics[width = 100mm, height = 100mm]{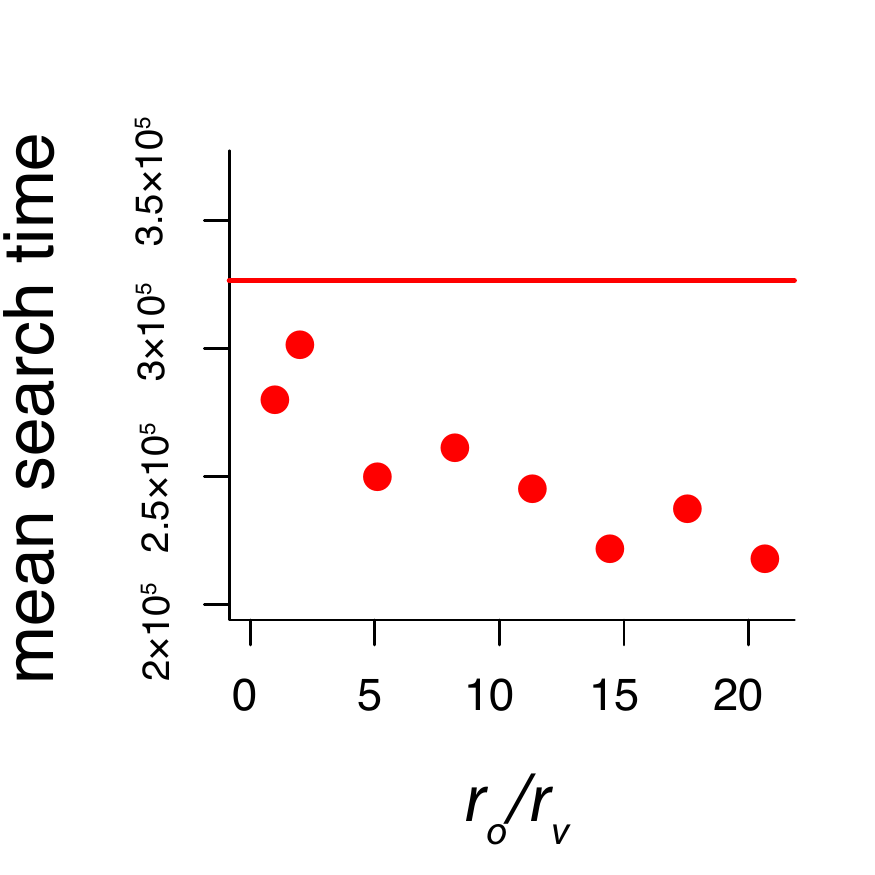}
\caption[Figure S4]{Mean search time of a signal-modulated L\'evy searcher that alters non-sensory behavior only when $h = 0$. Parameters as in Fig. \ref{fig:S1}. Each point represents mean of 200 replicate simulations.}
\label{fig:S4}
\end{figure}

To further explore the effect of no-signal events, we performed the following modification to the simulations described in the main text. We began with a predator that samples for olfactory signals during the scanning phase as the signal-modulated predators do. If the predator received a signal of $h > 0$, the next step length was drawn from a Pareto distribution as described for the non-sensory L\'evy strategy in the main text. However, when $h = 0$, the predator drew a step length from the distribution resulting from applying equation [1] in the main text, with $h = 0$. In other words, the predator behaved as a non-sensory L\'evy predator when $h > 0$ but as a signal-modulated L\'evy predator when $h = 0$. This is a convenient way to determine whether using no-signal events to exclude local regions of space is sufficient to improve search performance, or whether it is also necessary to use events where $h >0$. Results of this simulation show that altering behavior in response to no-signal events alone is sufficient to improve search performance at low target density (Figure \ref{fig:S4}). For example, when $r_o/r_v \approx 20$ predators that respond with signal-modulated behavior when $h = 0$ have mean search times that are 33\% shorter than mean search time of non-sensory L\'evy predators.

\bibliographystyle{pnas}

\end{document}